\documentclass[aps,prl,twocolumn,groupedaddress,amsfonts,amsmath,showpacs]{revtex4}
\usepackage{graphicx}
\usepackage{pdfpages}
\usepackage[applemac]{inputenc} 
\usepackage[T1]{fontenc}

\newcommand{\Lup}{{L_{\uparrow}}}
\newcommand{\Ldown}{{L_{\downarrow}}}
\newcommand{\Lp}{{L_{\rm p}}}

\begin{document}
\title{Hamiltonian traffic dynamics in microfluidic-loop networks}

\author{Raphaël Jeanneret} 
\author{Julien-Piera Vest}
\author{Denis Bartolo}

\affiliation{$^1$PMMH ESPCI-ParisTech, CNRS UMR 7636, Université Pierre et Marie Curie, Université Paris Diderot, 10 rue Vauquelin 75231 Paris cedex 05 France.}

\begin{abstract}
Recent microfluidic experiments revealed that large particles advected in a fluidic loop  display long-range hydrodynamic interactions. However, the consequences of such  couplings on the traffic dynamics in more complex  networks remain poorly understood. In this letter, we focus on the transport  of a finite number of particles in one-dimensional loop networks. By combining numerical, theoretical, and experimental efforts, we evidence that this collective process offers a unique example of Hamiltonian dynamics for hydrodynamically interacting particles.  In addition, we show that the asymptotic trajectories are necessarily reciprocal despite the microscopic traffic rules explicitly break the time reversal symmetry. We  exploit these two remarkable properties to account for the salient features of the  effective three-particle interaction  induced by the exploration of  fluidic loops.
\end{abstract}

%

\maketitle

The long-range nature of the hydrodynamic interactions is responsible for fascinating  collective phenomena in non-equilibrium suspensions, such as the velocity fluctuations of sedimenting particles~\cite{hinch2011}, and the emergence of  coherent structures in isotropic suspensions of active particles~\cite{saintillan2007}. 
However, in confined geometries, the walls screen exponentially the  correlations of the particle velocity~\cite{ liron}. Hence, no collective traffic phenomenon can occur when dilute suspensions flow in ducts having a width comparable to the particle size. Nonetheless, recent microfluidic experiments  in channels including a loop, revealed a rich variety of collective dynamics, such as multiperiodic and multistable traffic patterns
~\cite{jousse,whitesides,prakash,garsteckiloc,panizza1,panizza2,panizza3}. These experimental observations have been rationalized on the basis of  two empirical  rules~\cite{panizza1}:
as a particle enters a loop, it takes the branch in which the flow rate is maximal, and (ii) the particles partly obstruct the branch in which they flow.  
Consequently, the particle velocity at a node is a function of the  particle positions in the whole loop, thereby inducing localized but long-range hydrodynamic interactions.
So far, most of the research on  microfluidic traffic flows have been dedicated to the transport through a {\em single} fluidic loop fed at a constant rate by  a  continuous droplet/bubble stream.  

\begin{figure}
  \center
    \includegraphics[width=0.9\columnwidth]{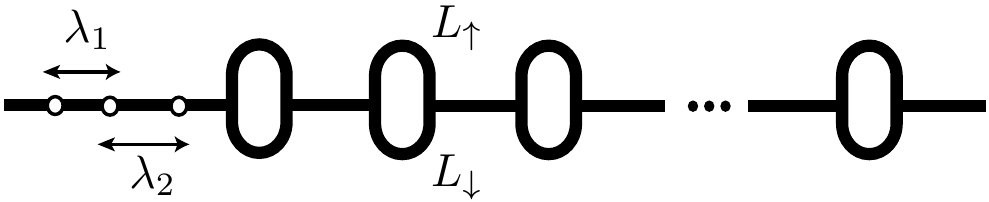}
         \caption{ Sketch of our theoretical and experimental set-up: three droplets are advected in a one dimensional microfluidic loop-network. $\lambda_2$ (resp $\lambda_1$) is the distance between the two righmost (resp. leftmost) droplets.}
    \label{fig1}
 \end{figure}

In this letter, we investigate the dynamics of a finite number of particles cruising in an extended  loop-network, see Fig.~\ref{fig1}.  We henceforth focus on the three-body problem. This setup is the basic building block to model the traffic dynamics of dilute suspensions (the case of two particles being trivial). Our primary idea is   to consider  the traffic through a single loop as a scattering process, which maps the distances $\lambda(n)=[\lambda_{1}(n),\lambda_{2}(n)]$ between the three particles entering  the loop $n$ into a new set of distances $\lambda(n+1)={\cal S}[\lambda(n)]$, where ${\cal S}$ is the scattering map. The transport through the entire network is then conceived as a discrete dynamical system, for which the loop index $n$ stands for the time variable.  From this perspective, we first evidence that, remarkably, the asymptotic traffic dynamics is Hamiltonian. To the best of our knowledge,  this is the only system of hydrodynamically interacting particles, for which an Hamiltonian description exists. Moreover, we show that the dynamics is asymptotically invariant upon time reversal symmetry despite the microscopic traffic rules are explicitly non-reciprocal. We exploit these two features to account for the geometrical and the dynamical properties of the scattering map $\cal S$. We close this paper, by comparing our theoretical predictions to microfluidic experiments. A quantitative agreement is found without any  free fitting parameter. 

We use a well established framework to model the traffic dynamics  in a fluidic network made of a chain of $N$ identical loops~\cite{jousse}. Precisely, it consists of four rules, which have proven to yield excellent agreement with the experiments~\cite{jousse,garsteckiloc,panizza1,panizza2,panizza3}: (i) The flow state of the fluid in the network is given by the analogous of the Kirchhoff laws. (ii) The particles are supposed to have a constant mobility coefficient. Therefore, we identify the fluid and the particle velocities.  (iii) When it reaches a vertex, a particle takes the branch where the fluid velocity is the higher. Note that this empirical rule, observed on deformable particles, is explicitly {\em non-reciprocal}. (iv) The particles partly obstruct the channels in which they journey. Precisely, the hydrodynamic resistances, expressed in unit-length, are given by: $L_{\downarrow,\uparrow}(n_{\downarrow,\uparrow})= L_{\downarrow,\uparrow}+n_{\downarrow,\uparrow}L_{\rm p}$,  where  the $n_{\downarrow,\uparrow}$  are the numbers of particles advected in the upper and in the lower branches respectively. The $L_{\downarrow,\uparrow}$ represent the branches' length, and $L_{\rm p}$ is the constant additional resistance induced by a single droplet. It follows that the particle velocity in the upper branch is:
\begin{equation}
v_{\uparrow}=v\,\frac{L_{\downarrow}(n_{\downarrow})}{L_{\downarrow}(n_{\downarrow})+L_{\uparrow}(n_\uparrow)},
\label{eq1}
\end{equation}
where $v$ is the fluid velocity outside the loops.
A symmetric formula holds for the lower branch.
 We apply  the above rules numerically using the  event-driven algorithm introduced in~\cite{schindler} for a 1-loop network and iterate it $N$ times. We record the distances $\lambda_{1}(n)$ and $\lambda_{2}(n)$, irrespective of the particle ordering, between the particles entering the loop $n$,  see Fig.~\ref{fig1}. The traffic dynamics is parametrized by two dimensionless numbers: the loop-aspect ratio $a\equiv \Lup/\Ldown$ and the "clogging-parameter" $c\equiv\Lp/\Ldown$, which quantifies how much a particle hinders the flow in a given branch.
In all that follows, we  restrain ourselves to weakly asymmetric loops for which $1<a<2$ and $a<1+c$. Within this approximation, a particle entering a loop journeys through the less occupied branch, or through the lower branch if the loop is empty.
 
  \begin{figure}
  \center
   \includegraphics[width=0.9\columnwidth]{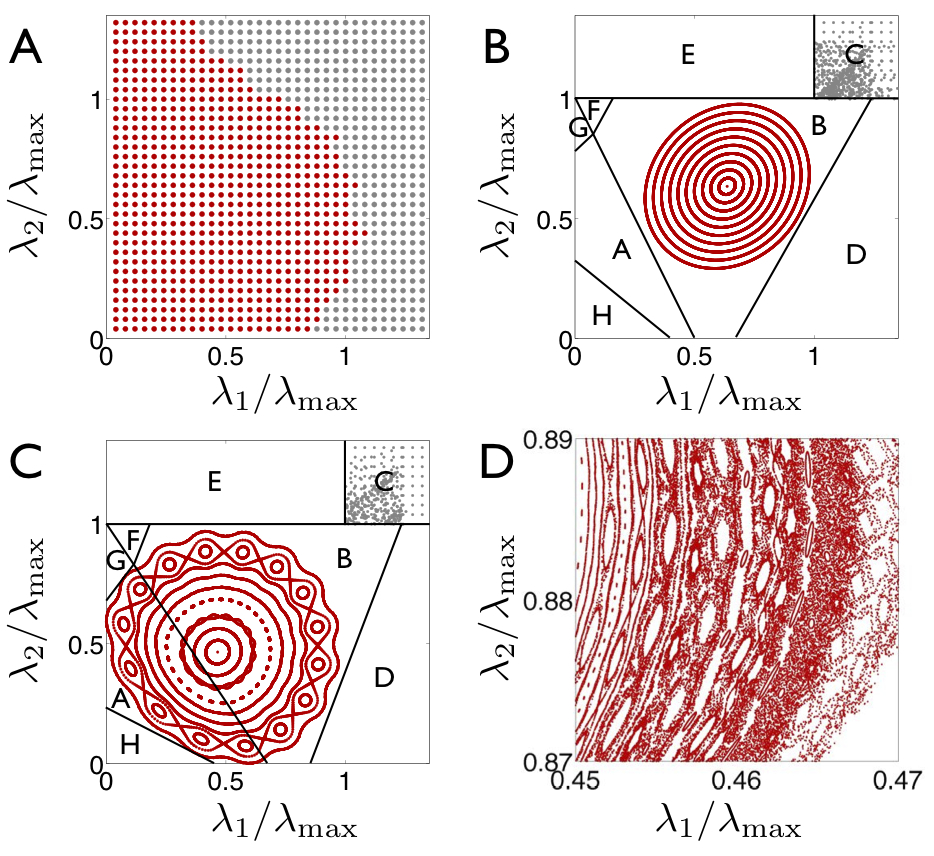}

         \caption{Numerical results obtained for: $a=10/9$. A-Basins of attraction of the $\cal S$-map.  The grey dots correspond to initial distances yielding stationary asymptotic dynamics (zone C).  The red dots converge to closed periodic orbits. $c=10/9$. B- Dots: Superimposed asymptotic trajectories for $c=20/9$. The 8 polygons correspond to the 8 trafficking scenarios. C- Dots: Superimposed asymptotic trajectories for $c=10/9$ (same parameters as in A). D- Close-up of  the edge of one island,  same parameters as in C.}
    \label{fig2}
 \end{figure}
The gross features of the traffic dynamics do not depend on the aspect ratio $a$. The phase plane $(\lambda_1,\lambda_2)$ is divided into two basins of attraction, Fig.~\ref{fig2}A. Starting from the rightmost basin, the system is quickly absorbed into the upper-right part of the plane, where the interparticle distances remain constant, Figs.~\ref{fig2}B and 2C. Starting from the leftmost basin, the system reaches a compact region filled with a continuous ensemble of periodic orbits centered on a unique fixed point. The distances oscillate around a constant value, and the orbits are either one dimensional curves, or zero dimensional (viz. returning repeatedly to a finite number of points). Two typical examples are given in Figs.~\ref{fig2}B and~\ref{fig2}C. These two regions are  reached after $\sim10$ loops.

\begin{figure}
  \center
    \includegraphics[width=0.8\columnwidth]{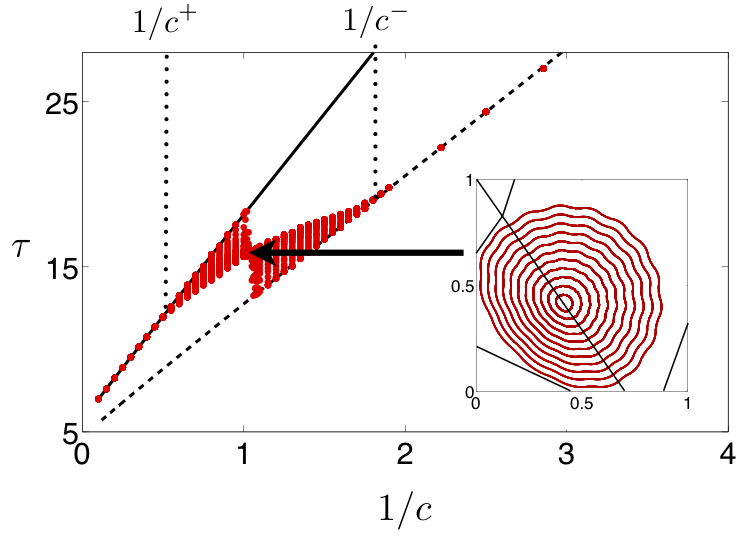}
         \caption{The period, $\tau$, defined from the maximum of the power spectra of $\lambda_1(n)$, is plotted versus the inverse of the clogging parameter $c$, for $a=10/9$. Each point corresponds to a different initial condition.  Full (resp. dotted) line: theoretical predictions for the periods  $\tau_B$ (resp. $\tau_A$). Inset: Phase portrait for $c=0.9628$, the period does not depend on the initial condition, but the self-similar trajectories are not elliptic. 
         }
    \label{fig3}
     \end{figure}

The absorbing region is defined by the inter-particle distance $\lambda_{\rm max}$ above which a  particle enters the loop after the previous one has left it. This distance is easily deduced from Eq.~\ref{eq1}: 
$\lambda_{\rm max}= \Ldown\left(1+\frac{1+c}{a} \right)$. Above $\lambda_{\rm max}$, the particles do not interact, and the scattering-map is trivial whatever $a$ and $c$: $\cal S={\mathbb I}$. 
Conversely, the geometry and the topology of the closed orbits strongly depend on the clogging parameter. First, we shall distinguish two regimes from the periods of the orbits. In Fig.~\ref{fig3}, we show the variations of the typical oscillation period, $\tau$, as a function of the clogging parameter $c$. There exist two limit values, $c^-$ and $c^+$, below and above which all the orbits share the same period. Moreover, for $c<c^-$, and $c>c^+$, the 1D-orbits are self-similar ellipses centered on a marginally stable fixed point, which lies on the line $\lambda_1=\lambda_2$, see Fig.~\ref{fig2}B. Therefore,  the $\cal S$-map is necessarily affine for this range of parameters. We emphasize that $\lambda_1=\lambda_2$ is a symmetry axis of the ellipses independently of $a$ and $c$. In contrast, for intermediate clogging parameters ($c^-<c<c^+$), $\tau$ strongly depends on the initial conditions,  and more than one period is detected, see Fig.~\ref{fig3}. In addition, at least one of the invariant curves has a non-elliptical shape. However, the global symmetry of the phase portrait with respect to  $\lambda_1=\lambda_2$ is preserved, see Fig.~\ref{fig2}C.  
Several closed orbits are  destabilized into  separatrix and island chains centered on stable $p$-periodic points. Trajectories with $p=15$ are clearly seen in Fig.~\ref{fig2}C. We systematically observed a hierarchy of island chains, as exemplified in  the close-up shown in Fig.~\ref{fig2}D. The inner part of the largest islands  clearly include island chains as well. They are separated by large chaotic regions, which also exist at the largest scale of the phase portrait, though they are much less extended.

We close this numerical section with the first main result of this letter. Remarkably, all the features of the  phase portrait are the hallmarks of Hamiltonian mappings, despite the traffic dynamics is a driven dissipative process. 
We shall note that fluid mechanics offers other examples of Hamiltonian descriptions for advected particles. However, these models  have so far been  restricted to non-interacting  passive tracers in  bidimensional and incompressible fluids, for which  the stream function readily provides an effective Hamiltonian~\cite{aref84}. The system, we consider here, does not belong to this class. Both the loop geometry and the effective hydrodynamic coupling between the particles, make impossible the use of a stream function as an effective Hamiltonian. 

\begin{figure}[t]
  \center
    \includegraphics[width=1\columnwidth]{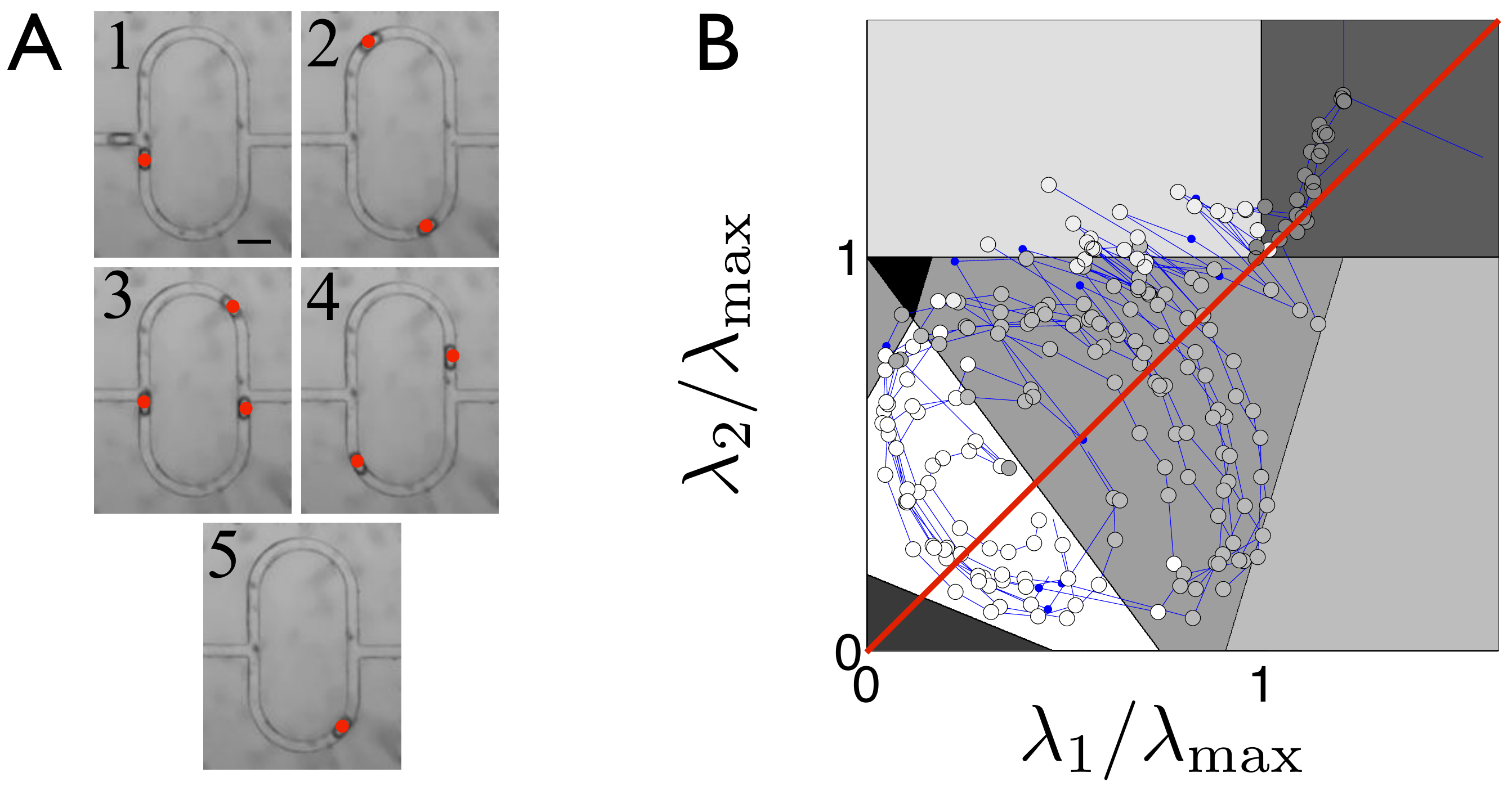}
         \caption{A: Five subsequent pictures of a typical experiment. Three drops (highlighted with a red dot) explore one loop according to the scenario $A$. B: Grey polygons: regions in which each scenario is expected from the theory. Connected dots: experimental trajectories recorded after $5$ loops have been explored. The greylevel of each dot codes for the observed trafficking scenario. The blue dots correspond to a loop including a geometrical defect.  Red line: $\lambda_1=\lambda_2$.}
    \label{fig4}
 \end{figure}

To  elucidate the Hamiltonian nature of the trafficking dynamics, we construct explicitly the scattering map $\cal S$. To do so, we first note that there exist 8 different traffic scenarios, labeled by $X=A,\ldots,H$. These scenarios are defined by the time ordered sequences of the  five occupation states, $(n_\uparrow,n_\downarrow)$, reached as the three particles journey through the loop. The system transit from one occupation state to  an other, when a particle reaches one of the two vertices of the loop. 
To make this definition clearer, we write explicitly the sequences corresponding to the two scenarios, which chiefly rule the asymptotic dynamics.  The scenario  $A=\lbrace(0,1),(1,1),(1,2),(1,1),(0,1) \rbrace$ is exemplified by the experimental pictures in Fig.~\ref{fig4}A. Three particles journey simultaneously in the loop, thereby inducing a change in the particle distances. Scenario $B=\lbrace(0,1),(1,1),(1,0),(1,1),(0,1) \rbrace$, the loop is explored at most by two particles simultaneously.  The other six traffic patterns are explicitly given, and sketched, in~\cite{supp}. 
Practically, $\cal S$ is a piecewise map, which has a different analytical expression, ${\cal S}_X$, for each  scenario. 
We first locate the regions of the phase plane in which each scenario prevails. To do so, using Eq.~\ref{eq1}, we compute the five times, $t_X^{(i)}$, $i=1\ldots5$, at which a particle reaches a vertex. The linearity of the  Kirchhoff laws, implies that the $t_X^{(i)}$ are linear functions of $\lambda_1$ and $\lambda_2$. Consequently, the region corresponding to the scenario $X$  is a polygon defined by the  inequalities: $t_X^{(i)}(\lambda_1,\lambda_2)<t_X^{(i+1)}(\lambda_1,\lambda_2)$.
The 8 polygons tile the phase plane as illustrated in Fig.~\ref{fig2}. We can then calculate the two distances  ${\cal S}_X(\lambda(n))=(\lambda_1(n+1), \lambda_2(n+1))$ by computing the time intervals, which separate the exit of two subsequent particles from the loop, and multiplying it by the fluid velocity outside the loops, $v=v_\uparrow+v_\downarrow$. Again, the Kirchhoff laws  require the ${\cal S}_X$ to be affine functions of the interparticle distances: $ \lambda(n+1)=M_X\cdot \lambda(n)+ L_X$, where  the $M_X$ and the $L_X$ are constant matrices and constant vectors. Their  exact  but lengthy expressions are given in the supplemental document~\cite{supp}. 

 We now exploit these analytical results to give a more physical insight on the geometrical and dynamical properties of the traffic dynamics.  Firstly, by superimposing the numerical trajectories on the eight regions of the phase plane, we notice that the asymptotic orbits are enclosed only in the union of the polygons $A$ and $B$, Fig.~\ref{fig2}. Moreover, the orbits that are enclosed in only one of those two regions  are ellipses.  To account for these observations, we compute the eigenvalues and the determinant of the $M_X$. Independently of the values of $a$ and $c$, $M_X$ is area preserving, $\det M_X=1$, in these two regions. Beyond our numerical observations,  this central result unambiguously proves  that the 3-particles dynamics is Hamiltonian in  $A$ and $B$. Furthermore, a tedious calculation proved that the  eigenvalues of $M_A$ and $M_B$ are two complex conjugate numbers, see~\cite{supp}. Consequently, the orbits are necessarily self-similar ellipses centered on a unique fixed point, when solely enclosed in $A$ or $B$, in agreement with our numerical results, Fig.~\ref{fig2}. 
 In addition, the system necessarily converges toward the three Hamiltonian regions $A$, $B$, and $C$ (region $C$ corresponds to the trivial case ${\cal S}_C=\mathbb I$). 
Indeed, $|\det M_X|$  takes only two different expressions elsewhere. $|\det M_X|=a(1+c)/(a+c)$, in regions $X=D,E,F,G$ and $|\det M_X|=(1+c)(a+c)/[a(1+2c)]$, in region $X=H$. In both cases we verify that $|\det M_X|>1$, as $1<a<1+c$. This implies that, asymptotically, the corresponding maps yield a continuous increase of $|\lambda_1|$ and $|\lambda_2|$.
 Therefore, as these maps are defined only in  polygons having a finite width, we conclude that  the system escapes from these regions as the particles flow through the loops. We also infer from this observation, that the largest invariant curve  is tangent to one of the boundary lines of the polygon $A\cup B$, see Fig.~\ref{fig2}. 

A second and important generic result is that the asymptotic traffic dynamics is time reversible. We now outline the demonstration of  this result, which we use to account for the symmetry of the phase portrait with respect to the $\lambda_1=\lambda_2$ direction. In this context,  time-reversal corresponds to the permutation of the inter-particle distances: ${\cal T}:(\lambda_1,\lambda_2)\to(\lambda_2,\lambda_1)$. Indeed, the last two particles  that exit a loop correspond to the first two entering particles when reversing the flow. Saying that $\cal S$ is time-reversible thus translates into ${\cal TSTS}={\mathbb I}$. This relation is obviously met along 1D trajectories enclosed in only one of the two regions $A$ or $B$. The corresponding traffic scenarios indeed correspond to palindromic sequences of occupation states. The same result can be also directly checked, by computing $({\cal T}{\cal S}_{X})^2$, where $X=A,B$,  using the analytic expressions of the affine maps given in~\cite{supp}. This identity is also satisfied  for trajectories overlapping the polygons $A$ and $B$ as well. The reason for this is that  ${\cal TS}_X(\lambda)\in X$ for the $\lambda$s belonging to the invariant curves of the region $X=A,B$. The demonstration of this last result  is tedious. It is detailed in the supplemental document~\cite{supp}. In order to show that the global symmetry of the phase portrait reflects the invariance upon time-reversal symmetry, let us consider a 1D orbit that crosses the symmetry line of $\cal T$, at a point  ${ \lambda}_{\rm s}=\cal T { \lambda}_{\rm s}$. Noting, that ${\cal T}^2=\mathbb I$ and ${{\cal S}^{-1}}=\cal TST$, we have $ {\cal S}^{-n}{\cal T}={\cal TS}^n {\cal T}$. Combining this relation and ${\cal S}^n\lambda_s={\cal S}^n{\cal T}\lambda_s$ yields ${\cal S}^n\lambda_s={\cal T}{\cal S}^{-n}\lambda_s$. This last identity precisely means that the entire orbit is  symmetric with respect to $\cal T$, as any $\lambda$ on this orbit can be generated from ${ \lambda}_{\rm s}$ ($\lambda={\cal S}^n{ \lambda}_{\rm s}$).

We now complete this description by a brief comment on the properties of the mean oscillation period, $\tau$, of the asymptotic dynamics. For small (resp. large) $c$, the orbits are included in the region $A$ (resp. $B$) only. Therefore, the periods $\tau_A$ and $\tau_B$ correspond to the argument of the eigenvalues of the matrices $M_A$ and $M_B$. They are plotted versus $1/c$ in Fig.~\ref{fig3} , using the analytic expressions given in ~\cite{supp}. 
For intermediate clogging parameters, the orbits overlap $A$ and $B$. Over a period, $n_A$ loops are explored according to the scenario $A$ and $n_B$ according to the scenario $B$. As the trajectories are closed curves around the fixed point, $\tau$ satisfies $2\pi/\tau=\left[2\pi\langle n_A\rangle/\tau_A+2\pi\langle n_B\rangle/\tau_B\right]/\left[\langle n_A\rangle+\langle n_B\rangle\right]$. This relation implies that the oscillation period is bounded by $ \tau_A$ and $ \tau_B$, in agreement with Fig.~\ref{fig3}. We also understand why there exists a unique period when the fixed point is on the boundary-line between $A$ and $B$.  As it includes the center of the ellipses, this line  separates the elliptic orbits of both regions into two identical parts. Therefore, $\langle n_A\rangle=\tau_ A/2$, and $\langle n_B\rangle=\tau_B/2$. This is again confirmed by the plots in Fig.~\ref{fig3} and Fig.~\ref{fig3} inset. 
 
Finally, to further confirm our theoretical predictions, we compare them to microfluidic experiments. Using the method introduced in~\cite{galas09}, we made a  device including $20$ identical loops ($\Lup=1,675\,\rm mm$, $\Ldown=1.525\,\rm mm$, channel width $75\,\mu\rm m$ and height $75\,\mu\rm m$). We monitored the trajectories of several triplets of identical water droplets advected by a continuous phase of hexadecane oil. By comparing the velocity of  an isolated droplet in the upper branches and in the straight parts of the channel, we deduce the experimental value of $L_{\rm d}=1.2\pm 0.25\, \rm mm$ from Eq.~\ref{eq1}. This  makes possible a direct comparison between our experimental and our theoretical results, without any free fitting parameter.  The evolution of  $\lambda_1(n)$ and $\lambda_2(n)$ are plotted in Fig.~\ref{fig4}B. The grey value of each point  
codes for the traffic scenario we observed experimentally. Though, the fine structure of the phase portrait cannot be probed with a 20-loops network, an excellent agreement between our experimental and theoretical results is found, when considering the three generic features of the asymptotic-dynamics: (i) The two asymptotic-dynamics schemes. The distances oscillate around a fixed point when $\lambda_1,\lambda_2<\lambda_{\rm max}$ and the traffic scenarios are of type $A$ or $B$ {\em only}. In contrast, when $\lambda_1,\lambda_2>\lambda_{\rm max}$,  we only observed small and   non-predictible variations of the $\lambda_i$. Complete freezing was never observed due to fluctuations in the droplet size, inducing differences in the droplets' mobility. (ii) Our model perfectly predicts  the location of the straight boundaries between the different traffic regions. (iii) The experimental phase portrait is  symmetric with respect to the $\lambda_1=\lambda_2$ direction.

In conclusion, combining experimental, numerical and theoretical tools,  we have provided a comprehensive description of the 3-body traffic dynamics. 
We expect that the generalization of our approach to coupled elementary traffic maps should provide a useful toolbox to design functional  microfluidic devices.

We thank Laurette Tuckerman,  Michael Schindler, Eric Lauga and Charles Baroud  for valuable comments and discussions. 
We acknowledge support by C'Nano IdF, Sesame Ile de France  and Paris \'emergence.


 \begin{figure*}
  \center
    \includegraphics[width=\textwidth]{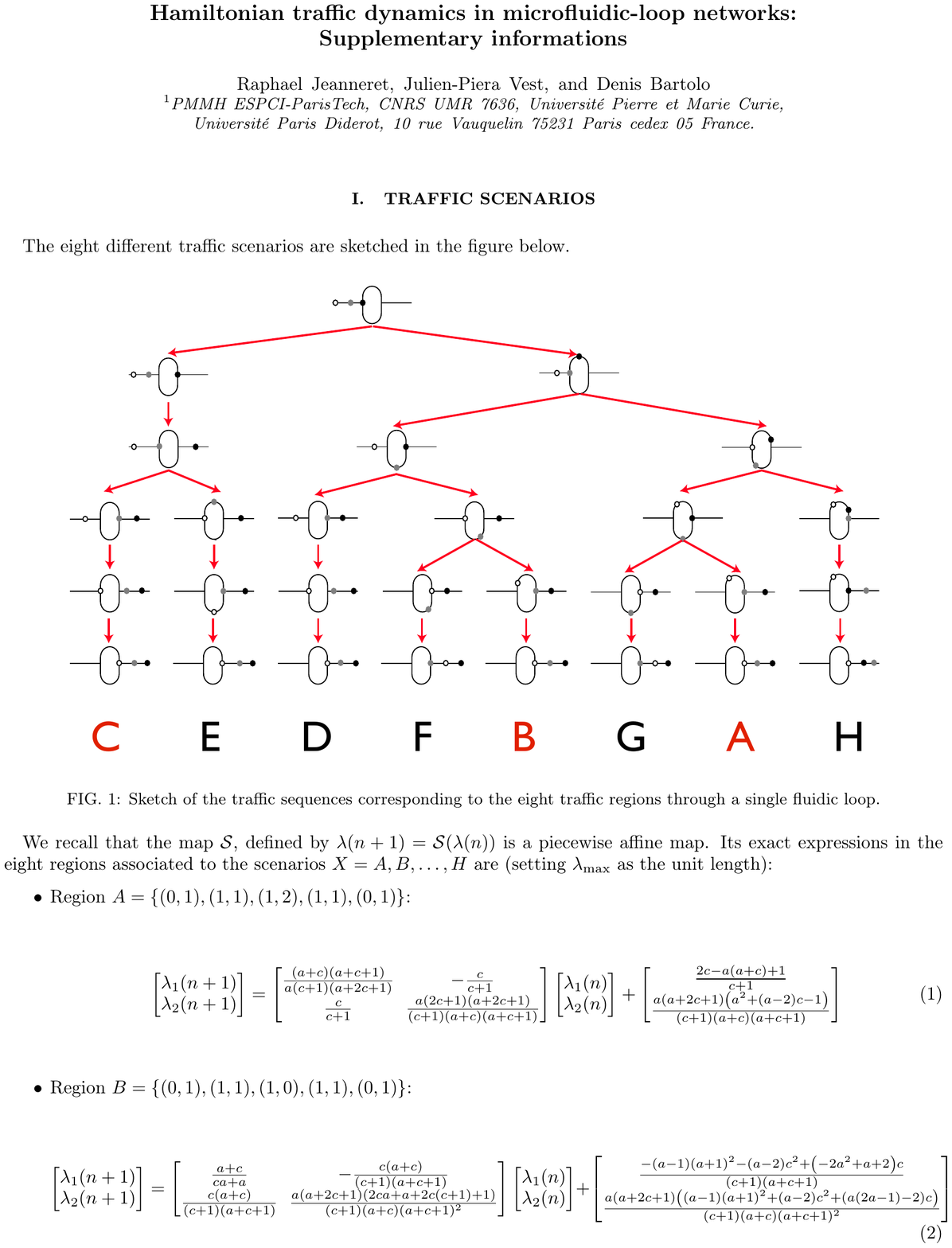}
   \end{figure*}
\begin{figure*}
  \center
    \includegraphics[width=\textwidth]{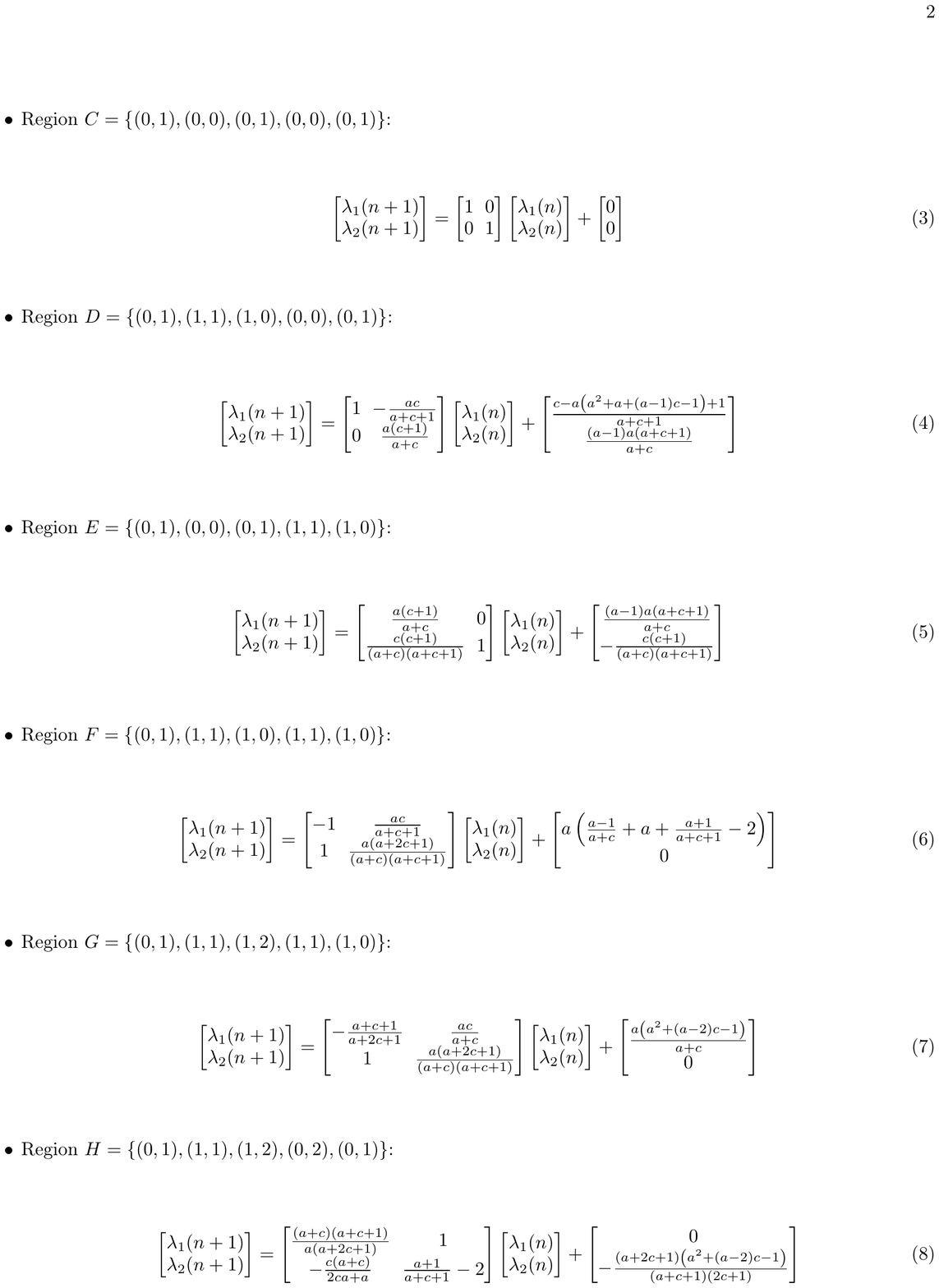}
   \end{figure*}
\begin{figure*}
  \center
    \includegraphics[width=\textwidth]{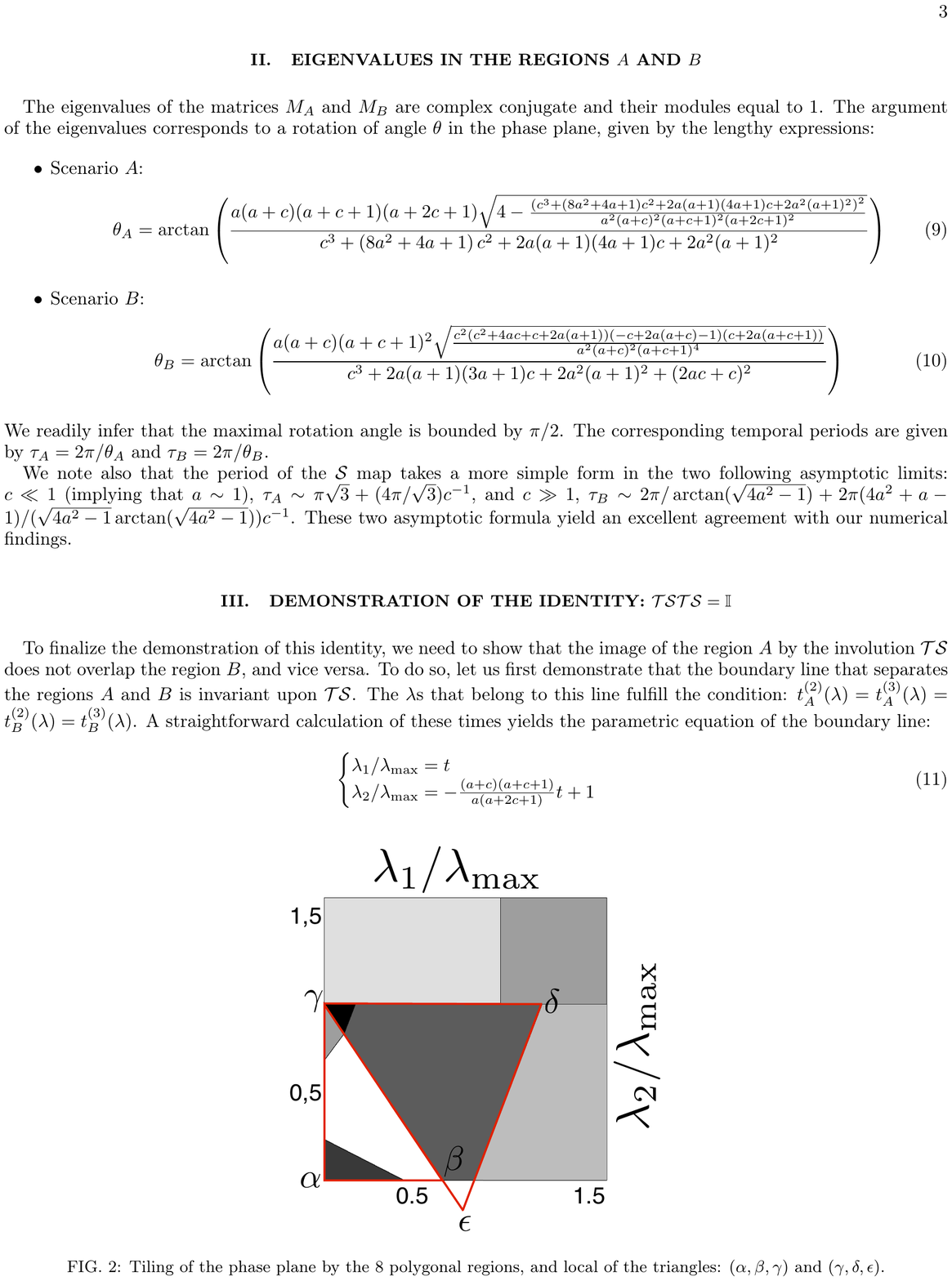}
   \end{figure*}
\begin{figure*}
  \center
    \includegraphics[width=\textwidth]{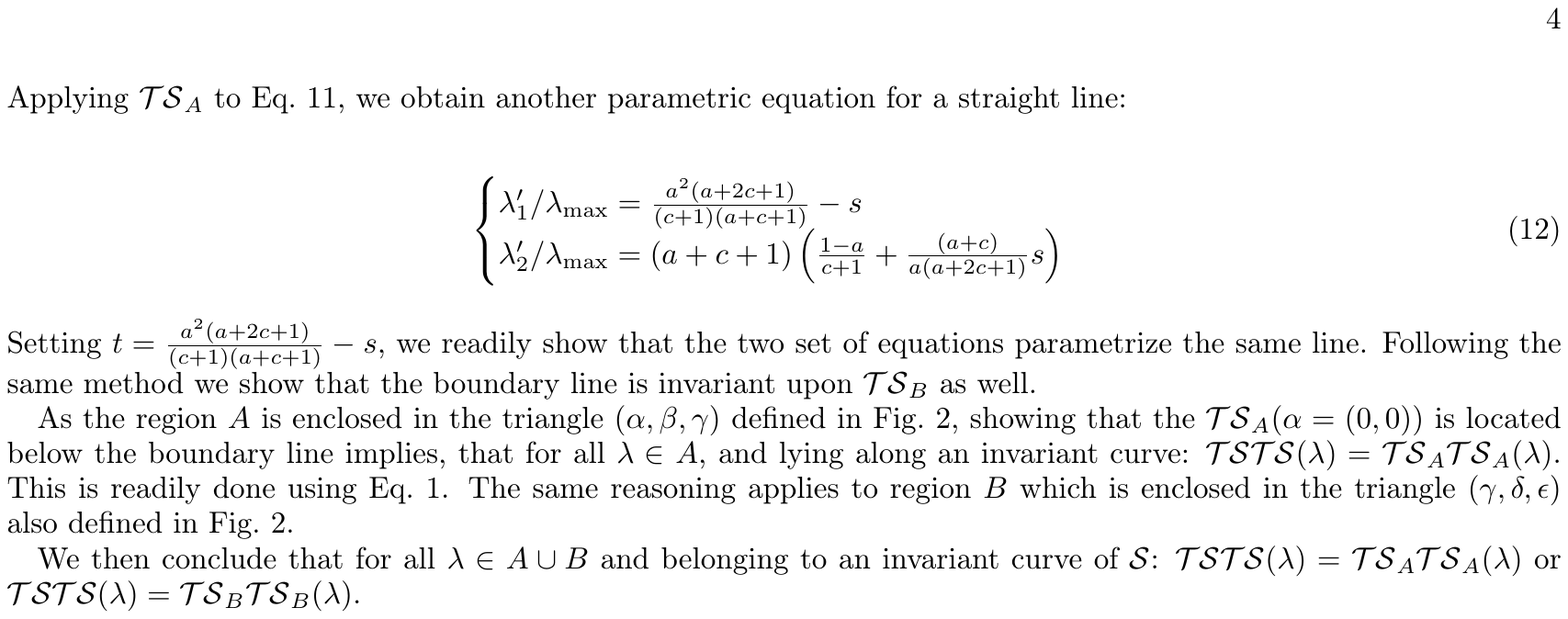}
   \end{figure*}

\end{document}